# Multi Scale Investigation of Surface Topography of Ball Python (Python Regius) Shed Skin in Comparison to Human skin

H. A. Abdel-Aal [1] M. El Mansori S. Mezghani
Arts et Métier ParisTech, Rue Saint Dominique BP 508,
51006 Chalons-en-Champagne,
France
[1] *corresponding author: Hisham.abdel-aal@chalons.ensam.fr*

**ABSTRACT**
A major concern in designing tribo-systems is to minimize friction, save energy, and to reduce wear. Conventional philosophy for design centers on mechanical and material considerations. In particular designers pay more attention to material properties and material choices based on mechanical properties rather than the design and shape of the contacting surfaces and the relation of that surface to the function of the device. As a result of thriving for miniaturization, focus has shifted toward optimal surface design (that is to construct a surface that is an integral part of the function of the tribosystem). Inspirations for such a trend come from studying natural systems and mimicking natural design rules. The major attraction is that natural systems, while functionally complex, are, in general, of optimized shape and performance. It is further believed that functional complexity of natural systems is what affords natural species to morph continuously to adapt with the operating environment. One bio-species that is of interest is the Ball Python. This is because such a species continuously slides against various surfaces, many of which are deemed tribologically hostile, without sustaining much damage. Much of the success of that species in adapting to its environment is attributed to surface design features. In that respect, studying these features and how do they contribute to the control of friction and wear is very attractive. This paper is a step in this direction. In this work we apply a multi scale surface characterization approach to study surface design features of the Python Regius. The focus is on those features that are typically used to assess the performance of high quality lubricating surfaces (such as those obtained through plateau honing). To this end, topographical features are studied by SEM and through White Light Interferrometery (WLI). We probe the roughness of the surface on multi scale and as a function of location within the body. In addition we draw a comparison of these features to those of human skin.

## 1. Introduction

In order for many next generation products to succeed they must offer greater functionality and improved performance. Ultra precision and structured surfaces are increasingly being adopted to gain such advantages. As such, the development of higher performing products through the application of ultra precise, complex and structured surfaces, such as those textured through plateau honing for improved lubrication capability, is an active area of research. In seeking inspirations for such custom designed surfaces many engineers turn toward natural systems (i.e., bio-species, plants, insects etc.,). There are many features that deserve attention within natural systems. These may include superior functionality, the ability to harness functional complexity to achieve optimal performance, and harmony between shape form and function. From a tribology perspective, the existence of surfaces that are design features, of the particular species, intended to facilitate functional performance is a point of deserving interest. Tribological investigations often deal with complex systems that, while nominally homogeneous, are practically compositionally heterogeneous. Compositional heterogeneity is either inherent (structural), or evolutionary (functional) [1]. Inherent heterogeneity is due to initial variation in composition, material selection, component chemistry, etc. Evolutionary heterogeneity, on the other hand, arises because of the evolution of the local response of different parts of a sliding assembly during operation. Subsystem components, for example, since they entertain different loads will react in a manner that is proportional to the local loading conditions. Distinct responses cause system subcomponents to evolve into entities that differ from their initial state. System heterogeneity, thus, introduces a level of functional complexity to the sliding assembly. Functional complexity, in turn, characterizes the interaction of system subcomponents, and of the system as a unit, with the surroundings. Most of such interaction, it is to be noted, takes place through the surface. Natural systems, regardless of the degree of functional complexity inherited within, display harmonious characteristics and an ability to self-regulate. Whence, as a general rule they operate at an optimal state marked by economy-of-effort. Man-engineered systems (MES), in contrast, do not exhibit such a level of optimized performance.

Much of the ability of natural systems to self regulate is attributed to optimized relationship between shape, form and function especially when surface design is considered. That is, shape and form in natural systems are always targeted toward optimal function. Such customization, however, is not advanced in MES. Bio-species, in that respect, offer many a lesson. This is because biological materials, through million years of existence, have evolved optimized topological features that enhance wear and friction resistance [2]. One species of remarkable tribological performance that may serve as an inspiration for optimal surface texturing is that of snakes. This is due to the Objective-targeted design features associated with their mode of legless locomotion.

Snakes are limbless animals. They have multi-modes of motion (slithering, crawling, serpentine movement etc) that take place during propulsion. Such motion modes are initiated through muscular activity. That is, through a sequence of contraction and relaxation of appropriate muscle groups. Transfer of motion between the body of the snake and the substrate depends on generation of sufficient tractions. This process, generation of traction and accommodation of motion, is handled through the skin. Thus the skin of the snake assumes the role of motion transfer and accommodation of energy consumed during the initiation of motion.

The number, type and sequence of muscular groups responsible for the initiation of motion, and thus-employed in propulsion, will vary according to the particular mode of motion to be initiated. It will also depend on the habitat and the surrounding environment. This also will affect the effort invested in initiation of motion and thereby also affects the function of the different parts of the skin and the amount of energy required to be accommodated. So that, in general, different parts of the skin will have different functional requirements. Moreover, the life habits of the particular species, e.g., defense, hunting, and swallowing) will require different deterministic functions of the different parts of the skin. That is the snake species, is a true representative of a heterogeneous tribo-system with a high degree of functional complexity, despite which, they don't suffer damaging levels of wear and tear.

Many researchers investigated the intriguing features of the serpentine family. Adam and Grace [3] studied the ultra structure of pit organ epidermis in Boid snakes to understand infrared sensing mechanisms. Johnna et al [4] investigated the permeability of shed skin of pythons (python molurus, Elaphe obsolete) to determine the suitability as a human skin analogue. Mechanical behavior of snake skin was also a subject of several studies as well. Jayne [5] examined the loading curves of six different species in uni-axial extension. His measurements revealed substantial variation in loads and maximum stiffness among samples from different dorsoventral regions within an individual and among homologous samples from different species. Rivera et al.[6] measured the mechanical properties of the integument of the common garter snake (Thomnophis sirtalis-Serpentine Colubridae). They examined mechanical properties of the skin along the body axis. Data collected revealed significant differences in mechanical properties among regions of the body. In particular, and consistent with the demands of macrophagy, it was found that the pre-pyloric skin is more compliant than post pyloric skin. Prompted by needs to design bio-inspired robots several researchers probed the frictional features of snake motion to understand the mechanisms responsible for regulating legless locomotion. Hazel et al [7] used AFM scanning to probe the nano-scale design features of three snake species. The studies of Hazel and Grace revealed the asymmetric features of the skin ornamentation to which both authors attributed frictional anisotropy.

In order to mimic the beneficial performance features of the skin, an engineer should be provided with parametric guidelines to aid with the objective-oriented design process. These should not only be dimensional. Rather, they should extend to include metrological parameters used to characterize tribological performance of surfaces within the MES domain. Thus, in order to deduce design rules there exists a need for quantification of the relationship governing micro-structure and strength topology of the bio-surface; exploring the quantitative regulation of macro and micro texture, and finally devising working formulae that describe (and potentially predict) load carrying capacity during locomotion in relation to geometrical configuration at both the micro and the macro scale. This paper is a preliminary step toward that goal.

In this work, we apply a multi-scale surface characterization approach to probe the design features of shed skins obtained from a Ball Python (Python Regius). Such a species is of tribological interest because of its locomotion taking place within a non-breakable boundary lubrication regime. Such a performance feature is facilitated by the topology of skin ornamentation. This renders such a species of interest to designers of plateau honing engineers where surfaces are designed for minimal lubricant consumption and for designers of hip and knee prostheses where maintaining a continuous boundary lubrication regime is a must. The emphasis in the current work, therefore, is on deducing those metrological aspects of the shed skin that are deemed essential to quantify the quality of tribological performance of industrial sliding assemblies (e.g., cylinder-piston).

## 2. Background
### 2.1. The Python species
Python Regius, figure 1-a, is a constrictor type non-venomous snake species that is typically found in Africa. Due to the reaction of the animal of curling into a spherical position, where the head is tucked under the trunk, figure 1-b,, for protection when distressed, it is also called "ball-python". The build of the snake is non-uniform as the head-neck region as well as that of the tail is thinner than the-region containing the trunk. The trunk meanwhile is the region of the body where most of the snake body mass is concentrated. Consequently it is more compact and thicker than other parts with the overall cross section of the body is more elliptical rather than circular. The tail section also is rather conical in shape. Skin of the python contains blotches imposed upon an otherwise black background. The blotches are of a non uniform shape and contour and are colored in dark and light brown. The ventral part of the body is typically cream, or extremely light yellow, color with occasional black spots scattered within.

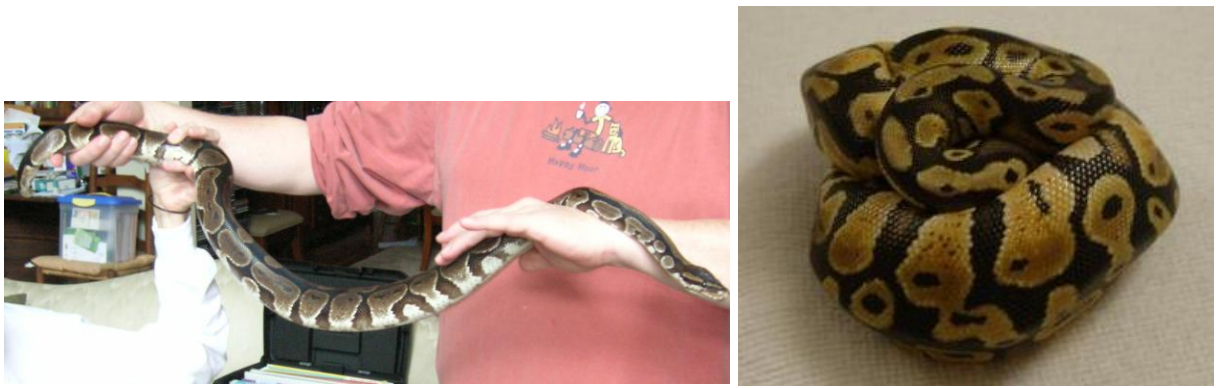

Figure 1. A Ball Python (Python Regius) species., a) normal position, b) curled-up "Ball" position

## 2.2. general features

The skin of reptiles, as well as that of snakes, is covered with scales. These maybe thought of as elevated folds or wrinkles within the skin. The pattern of the scales, shape, number, and ornamentation differ according to the species. A snake will typically hatch with a constant number of scales. This number remains fixed throughout the life duration of the snake. Each snake type has a different number of scales that is unique to the particular species. While the total number of scales remains constant, the size and shape change during the life span of the snake to accommodate growth and changes in size.

There are several functions for scales in a snake. They provide protection from the environment, they regulate energy exchange between the animal and environment, further they aid in camouflage and capture of prey [8]. Scales also perform a tribological function as they regulate frictional interaction, and help generate tractions upon propulsion. In snakes, skin cell division, occur periodically and that causes the reptile, to grow a second skin underneath the current (old) skin, and then "sheds" the old one. That is due to the cell division process the snake periodically sheds its entire outer skin layer.

### 3. Materials and Methods

All observations reported herein pertain to shed skin obtained from a 115 cm, 14 years old male Ball Python (Python Regius) housed individually in a glass container with news paper substrate. For optical microscopy observations skin was observed as is, whereas all samples for Scan Electron Microscopy observations were coated by a vacuum deposited Tungsten layer of thickness 10 nm. Surface topography analysis took place through two methods: SEM imaging in topography mode and through examination using a white light interferometer (WYKO NT3300 3D Automated Optical Profiling System). For comparison purposes, we obtained replicas of human skin. Samples were obtained by replicating the skin of a 40 years old female in two places: back of the hand and upper inside portion of the arm. All replicas were made using a light silicone rubber impression material (Silflo ®-Lexico, Plandent, UK) after using alcohol to clean subject skin.

## 4. Results and discussion

### 4.1. Initial Observations

Initial observations on the structure of the scales were performed using two methods: photography of the life species and optical microscopy. These were done without any treatment of the skin.

### 4.2. Photographical Observations

Figure 2 (a and b) depicts details of the surface structure of two regions on the life snake. Figure 2-a details the skin geometry in the head region from the inner (sliding side), whereas figure 2-b depicts the details of the stomach (belly-again the sliding side) of the snake. The photographs reveal that polygons constitute the geometrical building block of the surface. This polygon has eight sides, octagon, in the general area of the mouth (represented by the letter *O* in the figure). Past a line that joins the eyes, line AA, the pattern of the skin changes to hexagonal. The size of individual unit cells (octagons or hexagons) is not generally the same. The size of the octagons in the mouth region is not uniform. However, compared to the hexagonal patterns within the throat region, the area of the unit octagon is, in general, greater than that of the hexagonal unit. Hexagonal cells on the other hand are of uniform shape and size and seem to be of uniform density per unit area.

Within the mouth zone (above the line AA) the aspect ratio of the octagonal unit cell is, qualitatively, uniform. The major axis of the polygon, moreover, appears to point in the same direction of the body major longitudinal axis of the snake (BMLA). Considering the function of the head frontal part, feeding, some design features may be pointed out. Pythons feed by constricting the prey then swallowing it to be digested in the stomach. Upon inhaling the prey, the jaws have to stretch to accommodate the volume, and shape of the prey. The prey may offer some resistance that leads to multi-axial displacements of the surface material. Under such conditions, the design requirement of the surface is to allow for global flexibility and to facilitate local multi-axis displacements (stretching) all while minimizing possible damage to the skin. Observing the build of the skin above line AA, one would note that the skin is built of small patches of uniform shape (octagon), and are linked by a channel of what seems to be flexible strips of skin. Such a design allows for multi-axis stretching of the surface with possible damping of sudden jerks (caused by prey resistance) provided through the flexible links. A similar trend is noted in the throat region. Here the need to accommodate volume dilatation is greater since the prey by this point would, most likely, be subdued. Within this zone hexagonal unit cells make up the surface.

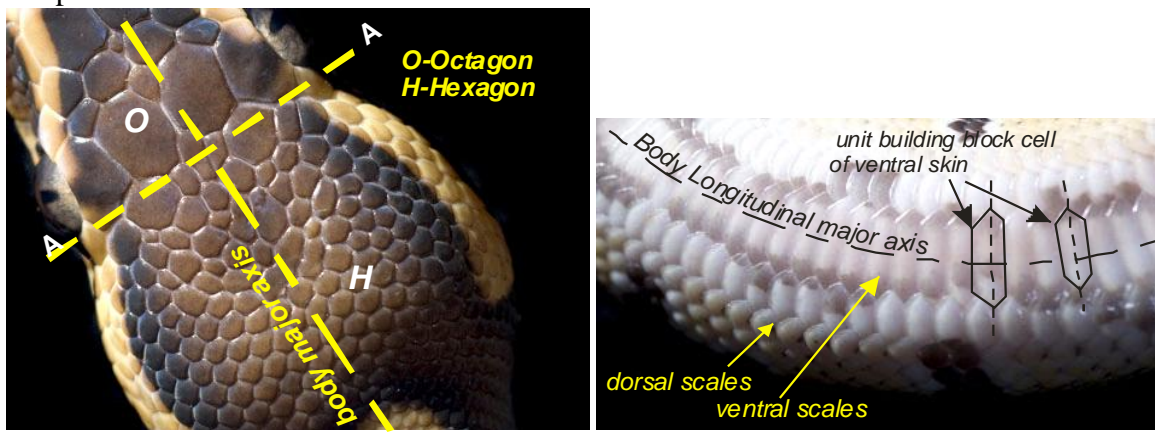

**Figure 2.** Details of scale structure at two positions on the life species.
 a) Shields within the head-throat region
 b) Ventral scales close to the waist section.

The uniform distribution of hexagonal cells is likely to aid the compliance of the surface and increase its flexibility. Interestingly such a hexagonal pattern is noted to be the most efficient way to pack the largest number of similar objects in a minimum space [9]. Naturally, had the skin been made of larger patches, or one continuous sheet, the probability of surface rupture would have increased. Figure 2-b depicts the surface geometry in the sliding side of the skin (general region of the belly). The photograph reveals that the hexagonal pattern constitutes the basic building block of the skin. The size of the hexagons differs around the circumference of the body. Large cells are particular to the main sliding area whereas cells of smaller size are particular to the back and the sides. Consistent with the construction of the head-throat region, the aspect ratio of the cells is variable. Of interest is the orientation of the major axis of the skin unit cells with respect to the BMLA. In the head-throat region the major axis of the cells is oriented parallel to the body major axis whereas in the belly (ventral scales) the major cell axis is perpendicular to the main body axis. Varenberg and Gorb [10], based on experiments on the hexagonal structures found on tarsal attachment pads of the bush cricket (*tettigonia viridissima*) suggest that variation in the aspect ratio of hexagonal structures may alter the friction force of elastomers by at least a factor of two. Additionally, we propose that the perpendicular orientation of the cells, with respect to the major axis of the snake, within the main sliding region aids in shifting the weight, and hence the contact angle and area of the snake upon sliding. Note that since the body of the snake is of cylindrical shape, the highest curvature of the skin will be oriented along the major cell axis. As such, upon sliding, the area of contact, and therefore the total tractions, will depend on the direction of motion (higher sideways and minimal forward). That is the orientation of the hexagon axis renders the friction forces isotropic. Such an observation is consistent with the findings of Zhang et al [11] who studied the frictional mechanism and anisotropy of Burmese python's ventral scales. They reported that the friction coefficient of the ventral scale had closely relationship with moving direction. The frictional coefficient for backward and lateral motion was one third higher than that in forward motion.

### 4.3. Structure of Shed Skin
#### 4.3.1. Optical Microscopy Observations

In snakes the epidermis is made up of different layers with the innermost called the stratum germinativum. The outer layers, which are renewed during shedding, are, from the inside, $\alpha$-, mesos-, β-layer, and *Oberhautchen*. The *Oberhautchen*, mainly consisting of β-keratin, is in direct contact with the environment and possesses a fine surface structure called micro-ornamentation [12]. Details of the micro-ornamentation were described by earlier authors [13-16]. Initial observations on the structure of the scales were performed using optical microscopy. without any treatment of the skin. Figure 3 depicts the structure of the scales at two positions within the skin in a region close to the waist of the snake. The first was from the back (dorsal scale) whereas the second position represented the belly of the snake (ventral). Note that although the general form of the cells is quite similar for both positions the size of a unit cell within the skin is quite different in both cases. In particular the cell is wider for the ventral (belly) positions. Each cell (scale) is also composed of a boundary and a membrane like structure. Note also the overlapping geometry of the skin and the scales (the so called scale and hinge structure). The skin from the inner surface hinges back and forms a free area, which overlaps the base of the next scale, which emerges below this scale figure 4.

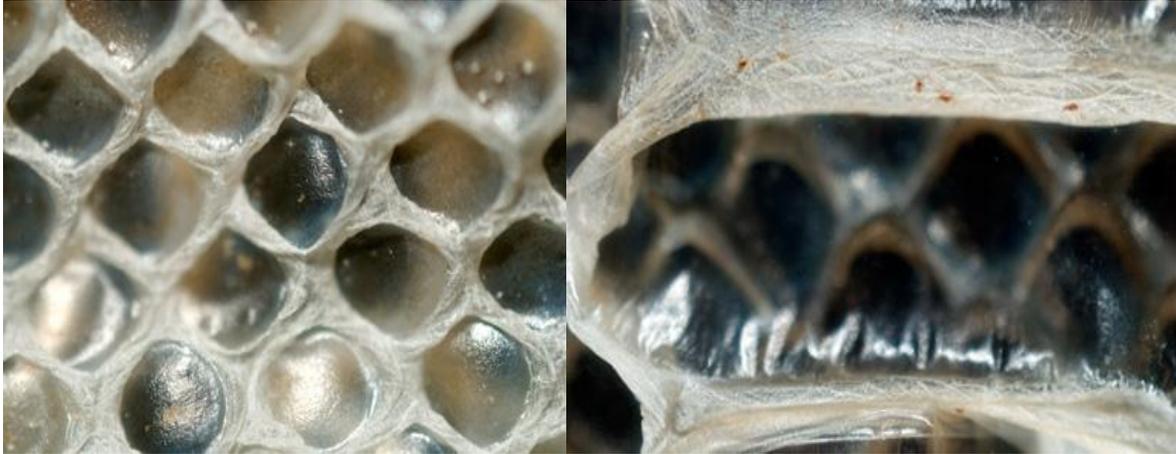

**Figure 3.** The structure of the scales on the inside of the shed skin at a region close to the mid section of the species at two orientations: back (dorsal) and abdominal (ventral).

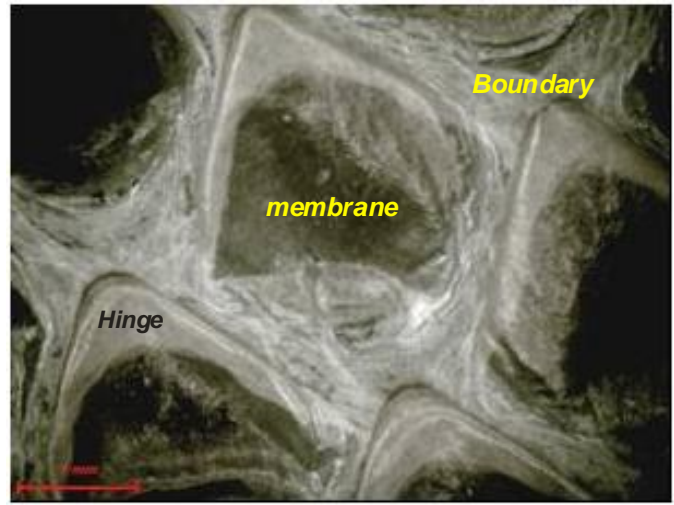

Figure 4. The details of dorsal scales from the inside of shed skin. The terminology used is: membrane to denote the major area of the scale and boundary to denote the raised part forming the circumference of the scale. Scale marker is 1mm.

### 4.3.2. Scan Electron Microscopy (SEM) Observations

Four positions on the shed skin were identified for initial examination. The choice of the positions was based on the functional profile of each position in the live species (figure 5). Position I is a representative of the neck region, position II represents the beginning of the trunk (waist) region, position III marks the boundary between the trunk and the tail region, finally position IV represents the tail region (containing the so called *subcaudal scales*). Skin swatches from each of the chosen positions were examined at different magnifications (X250-X15000) in topography mode. In order to suppress charging phenomena and improve the quality of observation, the surface of each sample was metalized by depositing a 10nm thin layer of platinum (Pt) using a sputter coater EMITECH K575X. Comparison of the uncoated and coated pictures is given in figure 6.

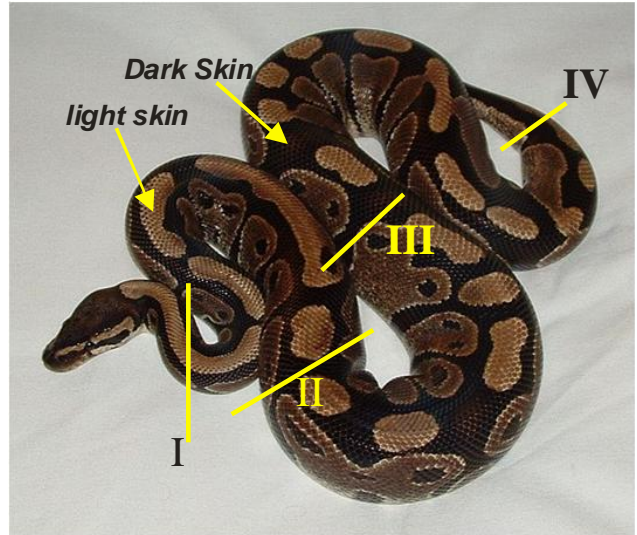

**Figure 5.** Equivalent positions chosen on the snake shed skin for SEM observations

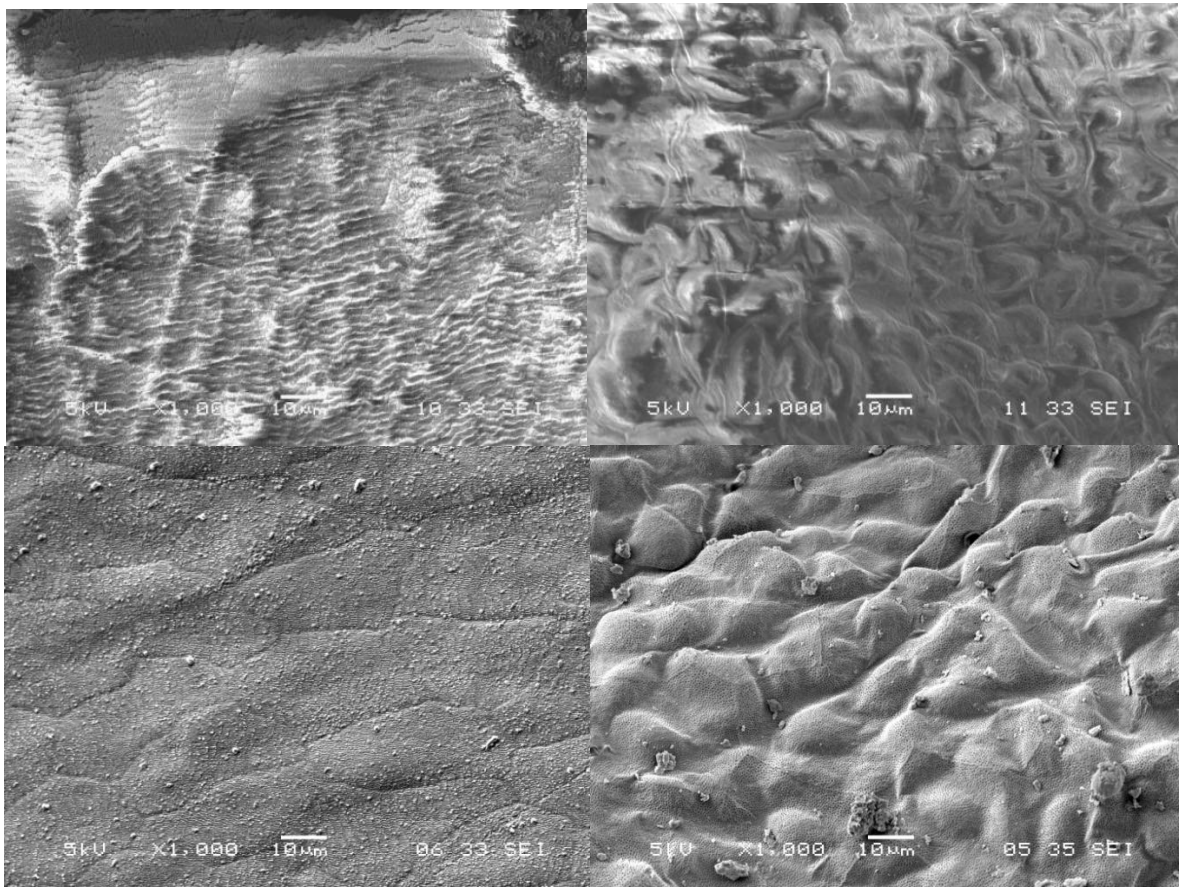

**Figure 6:** Comparison of the quality of SEM pictures at two positions on the shed skin at X1000. Upper row represents pictures of the uncoated skin samples while the lower row represents those of the coated samples.

For each position samples from the dark and the light colored skin (see figure 5) were also examined along with samples from the underside of the body. Major features of the observations are shown in figures 7 and 8.

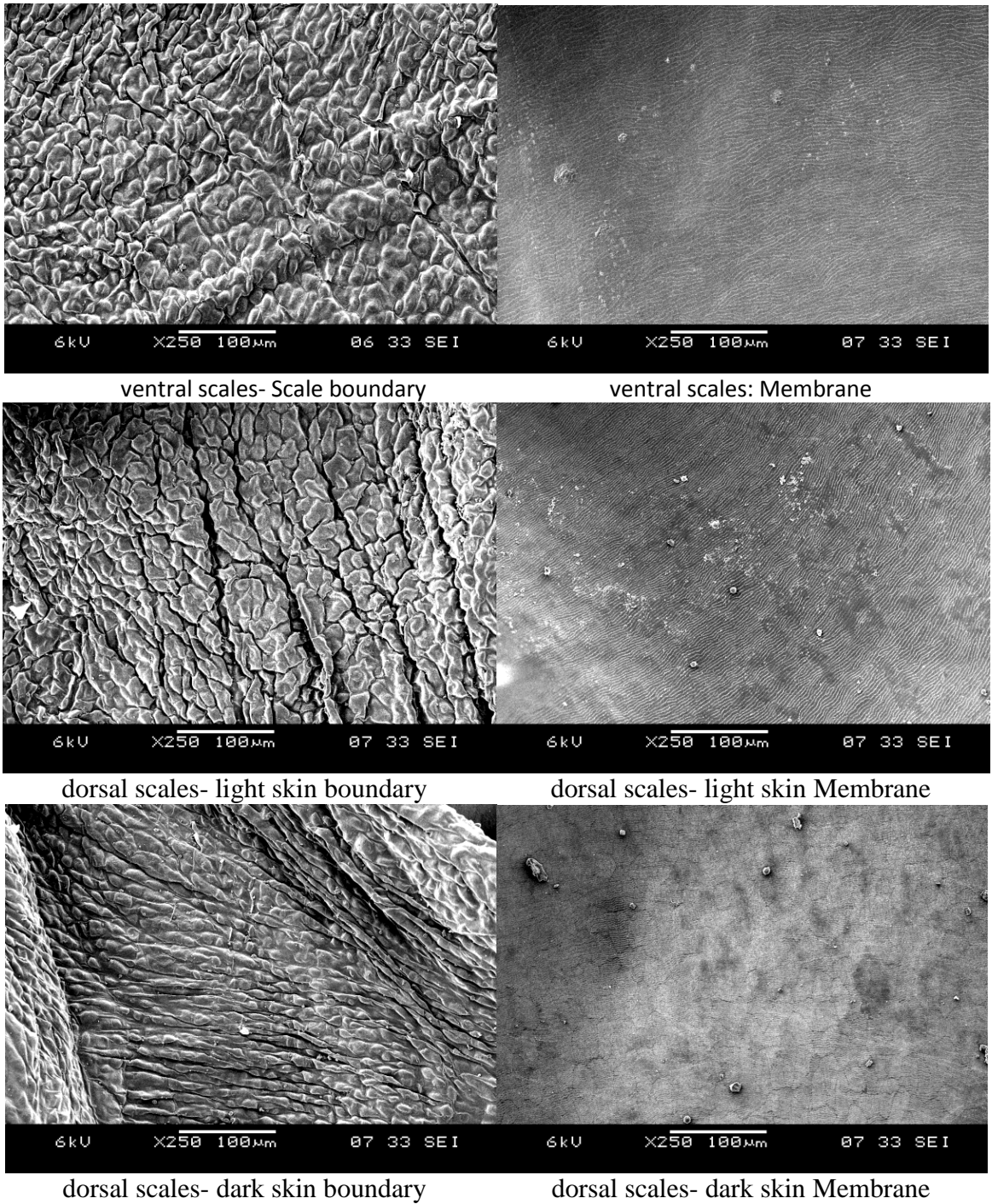

| ventral scales- Scale boundary | ventral scales: Membrane |
| dorsal scales- light skin boundary | dorsal scales- light skin Membrane |
| dorsal scales- dark skin boundary | dorsal scales- dark skin Membrane |

**Figure 7:** Major features of SEM observations of the skin swatches.(X-250)

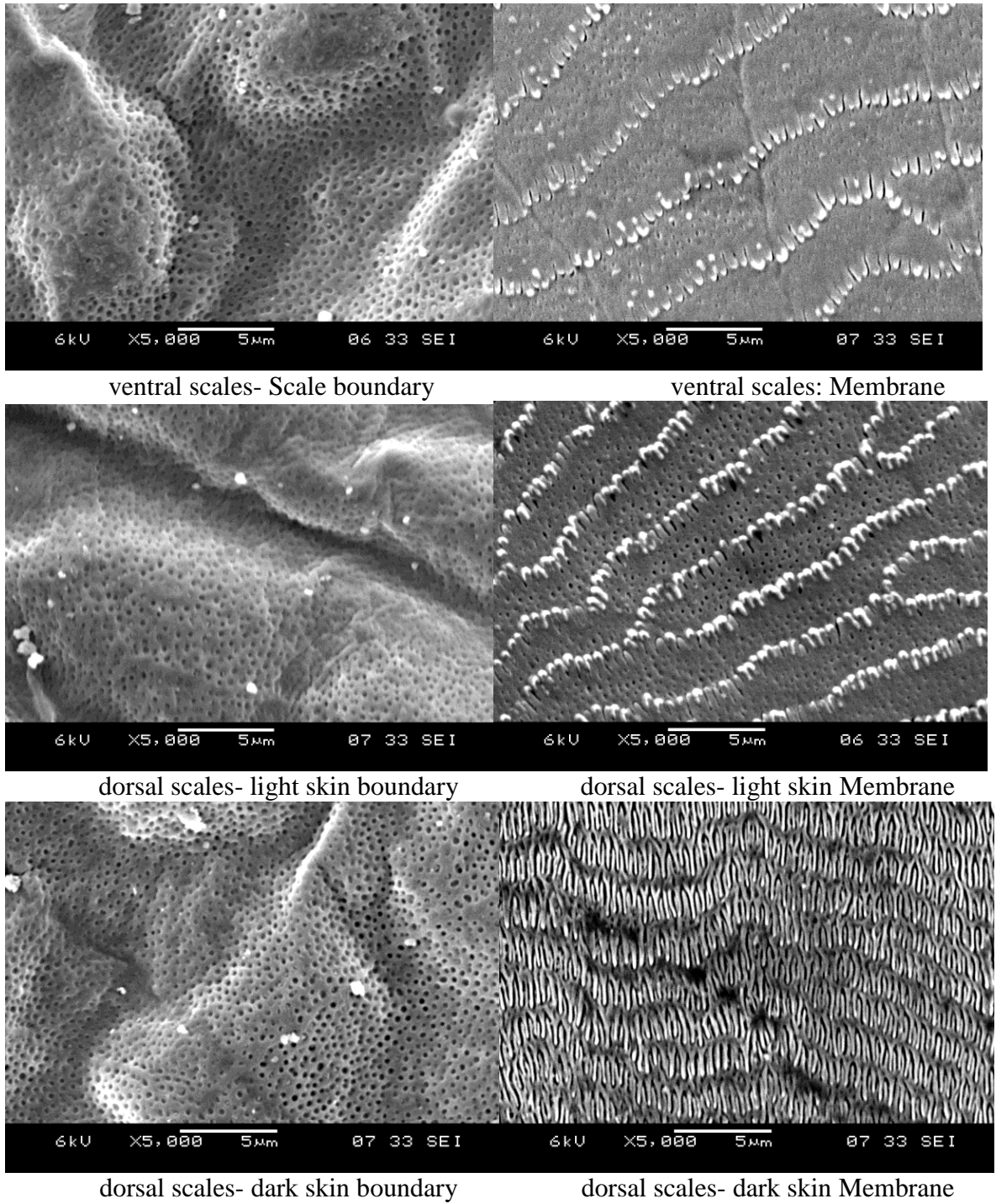

| | |
|---|---|
| ventral scales- Scale boundary | ventral scales: Membrane |
| dorsal scales- light skin boundary | dorsal scales- light skin Membrane |
| dorsal scales- dark skin boundary | dorsal scales- dark skin Membrane |

**Figure 8:** Major features of SEM observations of the skin swatches.(X-5000)

Note the inner structure that comprises pores. Two types of pores (or micro pits) may be distinguished: those located within the boundary and those located within the membrane. Image analysis of the pictures indicates that the diameter of the boundary-pores ranges between (200 nm – 250 nm). An estimate of the diameter of the membrane-pores was estimated by Hazel et al [7] using AFM analysis to be in the range (50nm -75 nm). Worth noting, is that within the same study [7] Hazel and co-workers scanned the contour of the fibril tips, within the membrane, using AFM and deduced that the tips are of asymmetric shape. Based on this observation, combined to the inter-lamellar layout of the fibrils, they suggested that such design features would lead to anisotropic frictional properties of the body. The locomotion mechanism responsible for such anisotropy is, however, considered, beyond the scope of the current work since the focus is on the topographical features and the metrology of the surface.

Surface protrusions are also noted within the boundary. These protrusions are of an asymmetric shape and irregular distribution. The surface of the membrane also comprises micro-nano fibrile structures. These are not of consistent shape and spacing. Note for example that the shape of fibril located in the dark colored skin region is different than that located within the light colored skin region (compare the X-5000 pictures). Moreover the density of the fibrils seems to be different within the different color regions (denser within the dark colored region). The spacing between different rows of fibrils also differs by skin color and position within the body (ventral> dorsal light skin> dorsal dark skin). Further analysis of images revealed that the density of the boundary-pores vary by position. That is the number of pores per unit area is not constant along the body, rather it changes relative to the position. Figure 9 is a plot of the variation in the density of the pores relative to the two sides of the skin (back-Dorsal scales and abdominal-Ventral scales) and in relation to the color of the skin (Light Patches (L) Vs Dark Patches D) within the back also.

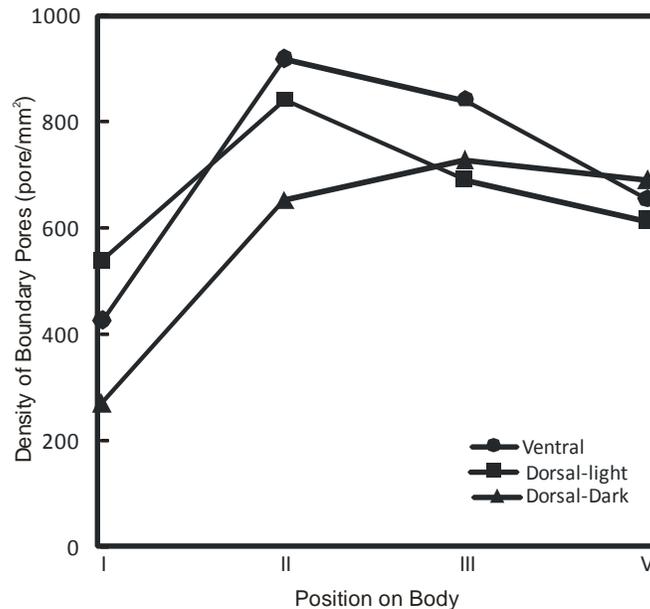

**Figure 9.** The variation in the density of the boundary pores (pore /mm$^2$) with position, and with color of skin.

### 4.4. Metrology of the Surface

Examination of the surface topography features of the skin using White Light Interferometry (WLI) on a swatches of skin (1500 μm x 1500 μm) yielded the basic parameters that describe the surface (asperity radii and curvature etc., ). Figure 10 depicts a typical WLI graph of the skin. The shown inteferogram pertains to a skin spot that is located along the waist of the snake from the belly side (ventral). Two interferograms are depicted: the one to the right hand side of the figure represents the topography of the cell- membrane whereas the one depicted to the left represents a multi-scale scan for the whole skin swatch. Note the scale on the right of the pictures as it indicates the deepest valley and highest point of the skin topography. For these typical skin swatches the value of the deepest part of the membrane was about -120 μm, whereas the highest summit is about 100 μm. The comparable values for the whole swatch are about -5.5 μm and 8.2 μm respectively.

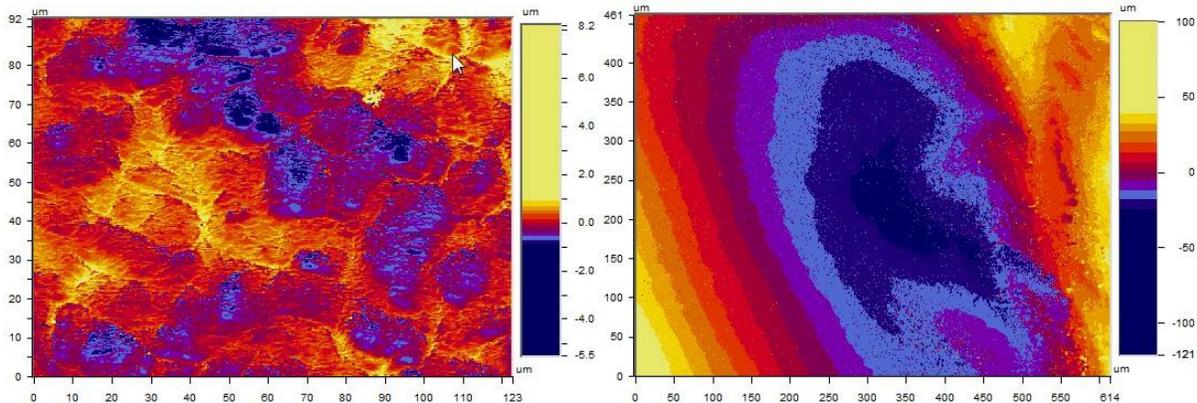

**Figure 10.** Multi scale WLI graphs depicting the topography of the skin building block (Scale) boundary and membrane (scale bar is in μm).

To establish a measure of comparison between the topography of the snake skin, we compared scans of snake skin to human skin scans. The results are given in figure 11.
Human scans from two different positions are depicted in the figure. The first scan is from skin located at the back of the hand whereas the second pertains to skin located at the inside of the upper arm of a 40 years old Caucasian female. Samples were obtained by replicating the skin using a replicating silicone. For the snake we chose to compare scans of the ventral scales located in the waist section of the body since it is a major load bearing area during locomotion. Again, it is noted that the peak summit and valley values of the surface of the snake skin are in the order of magnitude of one third that of human skin. Such a comparison highlights, qualitatively, the origins of the superior tribological performance exhibited by the snake. Note that under the conditions of examination in this work quantitative comparison between both skin types are not possible. This is because the tribological performance of human skin is not a mere function of surface topography. Rather, it also depends on the water content within skin cells. As such, comparing the frictional features of snake skin used in the current work, which is dry by default since it is shed from the species, to human skin in vivo is meaningless. Moreover, the friction of skin in general is dominated by acoustic emission. Again, provisions to monitor such effect were not undertaken in this study as it was judged that such feature warrants detailed investigation (which is currently undergone in our laboratory).

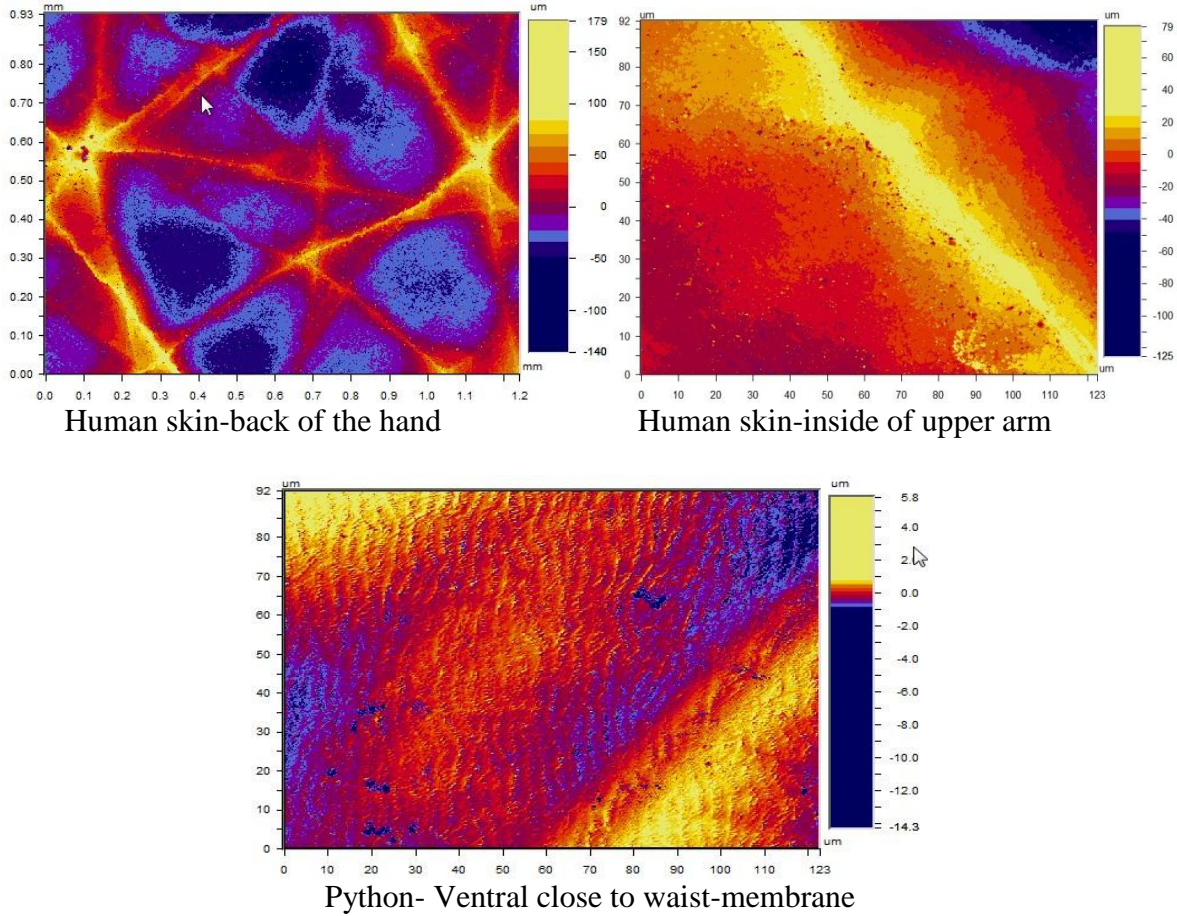

**Figure 11.** Comparison between the surface texture of snake skin and that of human skin replicas as revealed by White light interferometer (scale bar in μm)

### 4.5. Bearing curve analysis

To complete the analysis, we studied the load bearing characteristics of the skin at each of the predetermined positions (I through IV). Surface parameters were extracted from SEM topography photographs. The complete set of analyzed pictures provided a matrix of roughness parameters that describe the texture of the shed skin at variable scales ranging from X-100 to X-5000. Table 1 (a and b) provides a summary of the parameters extracted from the analysis. It can be seen that the scale of the analysis affects the value of the parameters, which may point at a fractal nature of the surface. Discussion of the implication of such finding is considered out of the scope of this work. However, of interest is to point out one of the features that directly relates to the design of the surface. Comparing the ratios between the Reduced Peak Height *Rpk*, Core Roughness Depth *Rk*, and Reduced Valley Depth *Rvk* reveals symmetry between the positions (compare the columns *Rpk/Rk*, *Rvk/Rk*, and *Rvk/Rpk* of table 1-b, and figure 12 a and b). This symmetry is interesting on the count that positions II and III represent the boundaries of the main load bearing regions (trunk). That is the regions on the body where the snake has most of his body weight concentrated (refer to figure 5) and thereby it is the region that is principally used in locomotion. Such symmetry may very well be related to the wear resistance ability of the surface or to the boundary lubrication quality of locomotion. Of interest also, is to find if implementing a surface of such characteristic parameters (functionally textured surface) in plateau honing for example would be conducive to an anti-scuffing and economical lubricant consumption performance. Such a point is a subject of ongoing investigation.

**Table-1 effect of magnification on surface parameters**

a- Surface parameters based on X-250 pictures

|              | Cr/Cf | Cl/Cf | Rpk/Rk | Rvk/Rk | Rvk/Rpk |
|--------------|-------|-------|--------|--------|---------|
| Position I   | 0.718 | 0.861 | 0.391  | 0.159  | 1.144   |
| Position II  | 2.011 | 2.010 | 0.612  | 0.545  | 0.656   |
| Position III | 2.066 | 1.628 | 0.733  | 0.436  | 0.621   |
| Position VI  | 1.388 | 0.926 | 0.617  | 0.195  | 0.749   |

b- Surface parameters based on X-5000 pictures

|              | Cr/Cf | Cl/Cf | Rpk/Rk | Rvk/Rk | Rvk/Rpk |
|--------------|-------|-------|--------|--------|---------|
| Position I   | 1.930 | 1.207 | 0.654  | 0.285  | 0.679   |
| Position II  | 1.273 | 1.803 | 0.478  | 0.404  | 0.812   |
| Position III | 1.671 | 1.622 | 0.484  | 0.359  | 0.800   |
| Position VI  | 1.772 | 1.158 | 0.636  | 0.260  | 0.688   |

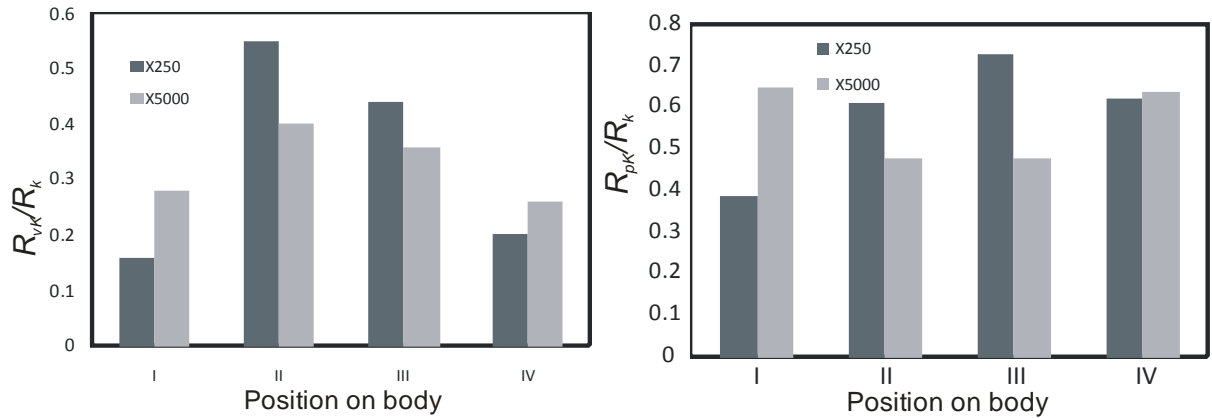

Figure 12: Plot of the ratio of the load bearing parameters Rvk/Rk and Rpk/Rk at two magnifications X-250 and X-5000.

## 5. Conclusions and Future Outlook

In this work we presented the results of an initial study to probe the geometric features of the skin of the Python Regius. It was found the structure of the unit cells is of regionally similar shape (octagonal and hexagonal).

Although almost identical in size and density, the skin constituents (pore density and essential size of the unit cell) vary by position on the body. Comparison of the topography of the snake skin to that of a human female revealed that the surface roughness of the snake species is around one third that of the human samples.

Analysis of the surface roughness parameters implied a multi-scale dependency of the parameters. This may point at a fractal nature of the surface a proposition that needs future verification. The analysis of bearing curve characteristics revealed symmetry between the front and back sections of the body.  It also revealed that the trunk region is bounded by two cross-sections of identical bearing curve ratios.  This has implications in design of textured surfaces that retain an unbreakable boundary lubrication quality and high wear resistance.

 Clearly much work is needed to further probe the essential features of the surface geometry. Namely, the basic parametric make up of the topography, form, and their relation to friction and wear resistance.

**Acknowledgment**

The authors acknowledge Mr. Benjamin Favre for assistance in SEM imaging and Mrs Ruth Ann Jones, of Troup county GA School System, for donating the shed skin used in this work.